\documentclass[%
 reprint,
 amsmath,amssymb,
 aps,
prl,
]{revtex4-2}

\usepackage{graphicx}
\usepackage{bm}
\usepackage{xcolor}
\usepackage{booktabs} 
\usepackage{siunitx}
\usepackage{amsfonts, amssymb, amsmath}
\usepackage{physics} 
\definecolor{green}{RGB}{0,180,0}

\newcommand{\gray}[1]{\textcolor{lightgray}{#1}}

\usepackage{comment}

\def\ec{{\rm e}} 

\def\Qlab{Q_{\rm lab}} 

\pdfoutput=1
\begin{document}

\title{Fine-structure constant sensitivity of the Th-229 nuclear clock transition}

\author{Kjeld Beeks$^{1,2}$}
\author{Georgy A. Kazakov$^{1}$}
\author{Fabian Schaden$^{1}$}
\author{Ira Morawetz$^{1}$}
\author{Luca Toscani De Col$^{1}$}
\author{Thomas Riebner$^{1}$}
\author{Michael Bartokos$^{1}$}
\author{Tomas Sikorsky$^{1}$}
\author{Thorsten Schumm$^{1}$}
\email{email: thorsten.schumm@tuwien.ac.at}
\affiliation{$^1$Vienna Center for Quantum Science and Technology, Atominstitut, TU Wien, 1020 Vienna, Austria}
\affiliation{$^2$Laboratory for Ultrafast Microscopy and Electron Scattering (LUMES), Institute of Physics, École Polytechnique Fédérale de Lausanne (EPFL), Lausanne CH-1015, Switzerland}

\author{Chuankun Zhang, Tian Ooi, Jacob S. Higgins, Jack F. Doyle, Jun Ye}
\affiliation{JILA, NIST and University of Colorado, Department of Physics, University of Colorado, Boulder, CO 80309}

\author{Marianna S. Safronova}
\affiliation{Department of Physics and Astronomy, University of Delaware, Newark, Delaware 19716, USA}
\date{\today}

\begin{abstract}\
State-resolved laser spectroscopy at the 10$^{-12}$ precision level recently reported in $arXiv$:2406.18719 determined the fractional change in nuclear quadrupole moment between the ground and isomeric state of $^{229}\rm{Th}$, $\Delta Q_0/Q_0$=1.791(2)\,\%. Assuming a prolate spheroid nucleus, this allows to quantify the sensitivity of the nuclear transition frequency to variations of the fine-structure constant $\alpha$ to $K=5900(2300)$, with the uncertainty dominated by the experimentally measured charge radius difference $\Delta \langle r^2 \rangle$ between the ground and isomeric state. This result indicates a three orders of magnitude enhancement over atomic clock schemes based on electron shell transitions.  We find that $\Delta Q_0$ is highly sensitive to tiny changes in the nuclear volume, thus the constant volume approximation cannot be used to accurately relate changes in $\langle r^2 \rangle$  and $Q_0$. The difference between the experimental and estimated values in $\Delta Q_0/Q_0$ raises a further question on the octupole contribution to the alpha-sensitivity.

\end{abstract}

\maketitle


\textit{Introduction.}--
The $^{229}$Th nucleus features a first excited metastable state, $^{229 \rm m}$Th, with an unusually low excitation energy of 8.4\,eV. This near-degeneracy of ground and isomeric state results from a fortuitous cancellation of the (repulsive) Coulomb and (attractive) nuclear energy contributions to the respective binding energies of the two states, which are each on the GeV level individually. 
This nucleus represents a unique opportunity to build a high-precision nuclear clock (see review~\cite{2021ThQST} and references therein), as the transition is very narrow, insensitive to external perturbations, and within the range of tabletop narrow linewidth lasers.

A nuclear clock based on trapped Th$^{3+}$ ions promises low systematic shifts, with an estimated total fractional inaccuracy of 10$^{-19}$~\cite{Campbell12}. Another unique opportunity enabled by the Th isomer is the development of a solid-state clock based on $^{229}$Th doped into a VUV-transparent crystal, where atoms are confined to a lattice in a space that is much smaller than the excitation wavelength realizing the Lamb-Dicke regime~\cite{Rellergert10,Kazakov12}. 

Resonant laser excitation of the isomer was first demonstrated in Th-doped CaF$_2$~\cite{2024Th} and then in Th-doped LiSrAlF$_6$ crystals~\cite{Elwell2024} using tabletop broadband tunable laser systems. Soon after, a frequency-stabilized VUV frequency comb was used to improve frequency resolution by $\sim$$10^6$ and directly resolve the nuclear quadrupole splitting in Th-doped CaF$_2$~\cite{Zhang24}. The nuclear transition frequency was determined to be 2~020~407~384~335(2)\,kHz. 

A variation of fundamental constants is predicted by many theories beyond the standard model of particle physics~\cite{SafBudDem18}. Moreover, ultralight dark matter may lead to oscillations of fundamental constants~\cite{Arvanitaki15,Flambaum3,Flambaum4,Antypas22, Kennedy2020}, and searches for such variation present a direct dark matter detection opportunity. If fundamental constants of nature change with space or time, so will atomic and nuclear energy levels, and consequently the associated clock frequencies. The amplitude of such changes however strongly depends on the particular clock transition under investigation~\cite{Flambaum1,Flambaum2,kozlov2018highly,safronova2019search,Faddeev20,barontini2022measuring}. Any spacetime-dependent variation of clock frequency ratios would be an unambiguous signal of new physics~\cite{Beloy2021, Filzinger2023}.

Highly enhanced sensitivity of the $^{229}$Th nuclear clock transition frequency to variation of the fine-structure constant $\alpha$ and the strong interaction coupling constant
$\alpha_S$ was predicted by Flambaum~\cite{Flambaum06}. Such an enhancement leads to increased discovery potential of any new physics that manifests as a variation of the fundamental constants~\cite{safronova2019search,Antypas22}.
However, the nuclear clock enhancement factor $K$ for the variation of $\alpha$ defined as $\frac{\delta \nu}{\nu} =K \frac{\delta \alpha}{\alpha}$ has yet to be precisely determined, where $\nu$ is the clock transition frequency. To determine $K=\Delta E_C/E$, where $E=h\nu$ is the isomer excitation energy, we need to know the difference in the Coulomb energy of the isomer and the ground state $\Delta E_C$~\cite{Flambaum06}. Since the Coulomb energies of both nuclear states are on the order of GeV, and $\Delta E_C$ is expected to be on the order of MeV, direct nuclear computation of the isomer and ground state energy is not sufficiently accurate for the determination of $K$ (see, for example~\cite{Flambaum09C}). Therefore, a method to determine  $\Delta E_C$ from measured nuclear properties of the ground and isomeric state using a geometric model was proposed~\cite{Flambaum09} but it hinges on a sufficiently precise measurement of the ratio of quadrupole moments of the two states. Hartree-Fock-Bogolubov nuclear many-body calculations~\cite{Flambaum09C} were used for validating the geometric model ~\cite{Flambaum09}. Here, we use the precision laser spectroscopy performed in~\cite{Zhang24} to extract the nuclear quadrupole splitting and determine the sensitivity of the nuclear clock to $\alpha$-variation.

\textit{Quadrupole structure.}--
In the nuclear laser spectroscopy performed in~\cite{Zhang24}, samples containing $^{229}$Th nuclei doped in a CaF$_2$ matrix were used~\cite{beeks2022nuclear,beeks2023growth,beeks2024optical}.
The interaction of the nuclear quadrupole moment ($Q_0$) in the lab frame ($Q_{\rm lab}$) with the crystal electric field gradient (EFG), $V_{ij}={\partial^2V}/({\partial x_i \partial x_j})$, leads to the emergence of a nuclear quadrupole splitting~\cite{Kazakov12}.

The intrinsic quadrupole moment $Q_0$ in the body-fixed frame is connected with the spectroscopic quadrupole moment $Q_{\rm lab}$ in the lab frame via the relation
\begin{equation}
Q_0=\frac{\Qlab}{q_e Z}\frac{(2I+3)(I+1)}{3k^2-I(I+1)},
\label{eq:QviaQ0}
\end{equation}
where $Z=90$ is the charge number of the thorium nucleus, $q_e$ is elementary charge,  $I$ is the nuclear spin ($I=5/2$ for ground, and $I=3/2$ for isomeric state), and $k$ is the projection of the nuclear spin on the deformation axis. To determine the enhancement factor, $K$, we rely on the ratio of quadrupole moments of the nuclear isomer and ground state $Q_0^\mathrm{m}/Q_0$ from the experimentally determined nuclear electric quadrupole splitting, see Ref.~\cite{Zhang24}.


We first describe the interaction Hamiltonian between the EFG $V_{ij}$ and the spectroscopic nuclear quadrupole moment $\Qlab$ and derive the energy levels and eigenstates. Here we define axes of the EFG such that $V_{ij}$ is diagonal, $|V_{\rm zz}|$ is the maximum value of the EFG and $\eta=(V_{\rm xx}-V_{\rm yy})/(V_{\rm zz}$) is the asymmetry of the EFG. We then calculate the energies of the quadrupole states and excitation probabilities $W(E_2-E_1)$ by calculating the transition matrix elements of the interaction Hamiltonian between the eigenstates and the driving laser field. The results of these calculations can be found in Table~\ref{table:lines}. 
A detailed derivation can be found in the supplementary material. 

\begin{table}[htbp]
\centering
\caption{Observed lines and their interpretations according to~\cite{Zhang24}, together with calculated relative intensities $W$.} 
\label{tab:crystals}
\begin{tabular}{@{}llccc@{}}
\toprule
\textbf{$j$}\hphantom{aaa} & $\nu_j$ [THz] & $~m_g$ & $m_e$ & $W(E_2-E_1)$ \\ \midrule
1 & 2020.407 283 847(4)\hphantom{a} & \hphantom{a} $\pm 3/2$ \hphantom{a} & \hphantom{a} $\pm 1/2$ \hphantom{a}  & 0.184\\
2 & 2020.407 298 727(4) & $\pm 5/2 $ & $\pm 3/2$ & 0.328\\
3 & 2020.407 446 895(4) & $\pm 1/2 $ & $\pm 1/2$ & 0.310\\
4 & 2020.407 530 918(4) & $\pm 3/2 $ & $\pm 3/2$ & 0.149\\
5 & 2020.407 693 98(2) & $\pm 1/2 $ & $\pm 3/2$ & 0.0233\\
\gray{6} & \gray{2020.407 051 655}\footnote{unobserved, asymptotically forbidden transition}   & \gray{$\pm 5/2 $} & \gray{$\pm 1/2$} & \gray{0.0058}\\
\bottomrule
\end{tabular}
\label{table:lines}
\end{table}


\textit{Data Fitting.}--
In Ref.~\cite{Zhang24}, an absolute frequency measurement of 5 lines was performed, which were identified as transitions between certain quadrupole sublevels of the ground and the isomeric state of the $^{229}$Th nucleus, see Table~\ref{table:lines}. The transition energies were fit to analytical solutions of the Hamiltonian to yield $\eta$, $\Qlab V_{\rm zz}$ and $\Qlab^\mathrm{m} V_{\rm zz}$ thus obtaining the ratio of $\Qlab^\mathrm{m}/\Qlab$. The frequency of the EFG-free isomer transition $\nu_{\rm Th}$ was determined by a weighted averaging of the line frequencies.

Here, we independently verify the data fitting using the numerical solutions of the interaction Hamiltonian with $\Qlab V_{\rm zz}$, $\eta$, $Q_0^\mathrm{m}/Q_0$ and $\nu_{\rm Th}$ as free parameters. Details of the employed statistical methods and in particular the estimation of uncertainties can be found in the supplementary materials. We obtain $\nu_{\rm Th}=2020\,407\,384\, 335(2)~{\rm kHz}$, 
$Q_0^\mathrm{m}/Q_0=1.01791(2)$, $\eta=0.59164(5)$ and
$\Qlab V_{\rm zz}=339.263(7)~{\rm eb\, V/\r{A}^2}$. The fit results are fully consistent with the results reported in Ref.~\cite{Zhang24}.

The experimental line intensities reported in~\cite{Zhang24} qualitatively agree with the theoretical expectations. The transition frequencies are arithmetically consistent and stable over the measurement campaign; we therefore assume fluctuations $V_{ij}$, i.e. due to changes in temperature or pressure, to be negligible. The temperature of the crystal in the measurement was 150(1)\,K, a detailed study of the above parameters with crystal temperature will be a topic of further studies. 

Using the value $\Qlab=3.11(2)$\,eb for the $^{229}$Th ground state quadrupole moment derived in~\cite{2021Th3} yields $V_{\rm zz}=109.1(7)\,{\rm V/\r{A}^2}$. The rather high asymmetry parameter $\eta$ indicates a non-rotationally symmetric microscopic charge compensation configuration involving multiple atoms, such as two fluoride interstitials~\cite{dessovic2014229thorium,pimon2022ab}. The microscopic atomic and electronic structure of the $^{229}$Th defect in the CaF$_2$ crystal will be subject of an upcoming publication. Here we focus on the consequences of the fractional change of $Q_0^\mathrm{m}/Q_0$ for the nuclear properties of $^{229}$Th.

\textit{Nuclear Theory.}--
Following~\cite{Flambaum09,Thielking18,Faddeev20}, we describe the nucleus as a prolate spheroid with uniform volume charge. The mean-square radius $\langle r^2 \rangle $ and the intrinsic quadrupole moment
$Q_0$ are expressed via semi-minor and semi-major axes $a$ and $c$ as
\begin{align}
\langle r^2 \rangle = \frac{1}{5}\left( 2a^2+c^2\right), ~~~~~
Q_0=\frac{2}{5}\left(c^2 - a^2\right). 
\end{align}

Very coarsely, upon nuclear excitation, the distribution of a valence neutron is modified and polarizes the proton distribution via the strong interaction, altering the mean-square charge radius $\langle r^2 \rangle$ as well as the nuclear deformation and in consequence the electric quadrupole moment $Q_0$. These two quantities, therefore, are essential for the determination of the Coulomb energies of the nuclear ground and isomeric state.
The Coulomb energy~\cite{Beringer61} is given by Eq.~(5.11) in Ref.~\cite{nucl},
\begin{equation}
E_C=\frac{3q_e^2 Z^2}{5R_0} \frac{(1-\ec^2)^{-1/3}}{2\ec}{\rm ln}\frac{1+\ec}{1-\ec},
\end{equation}
where $\ec^2=1-(a^2)/(c^2)$ 
is the eccentricity and $R_0^3 =a^2c$
is the equivalent sharp spherical radius, see Eq.~(\ref{eq:21}).
The difference of the Coulomb energy of the isomer and the ground state is computed as
\begin{eqnarray}
\Delta E_C &=&\langle r^2 \rangle \frac{\partial E_C}{\partial \langle r^2 \rangle} \frac{\Delta \langle r^2 \rangle}{\langle r^2 \rangle} +
Q_0  \frac{\partial E_C}{\partial Q_0} \frac{\Delta Q_0}{Q_0}\\
&=&-485\ \textrm{MeV} \frac{\Delta \langle r^2 \rangle}{\langle r^2 \rangle} + 11.3\ \textrm{MeV} \left( \frac{Q_0^\mathrm{m}}{Q_0}-1
 \right), \label{ec}
\end{eqnarray}
where we used $\langle r^2 \rangle=5.756(14)$\,fm$^2$~\cite{Angeli13}, and the spectroscopic quadrupole moment
$\Qlab=3.11(2)$\,eb~\cite{2021Th3} for the ground state. The ground state spectroscopic quadrupole moment $Q_{\rm lab}$ was recently extracted from experimental measurements of the electronic hyperfine-structure~\cite{Campbell11} using the coupled cluster method with single, double, and triple excitations for both valence and core electrons, improving the accuracy by a factor of 3~\cite{2021Th3}. This newer value accounts for a small difference in the second coefficient 11.3~MeV compared to~\cite{Thielking18,Faddeev20}. We express Eq.~(\ref{ec}) in terms of the difference of $\langle r^2 \rangle$ and the ratio of quadrupole moments as these are the quantities which are determined from the experiments, see discussion below.

\textit{Results and Discussions.}--
We start with the discussion of the experimental results used to compute the $\Delta E_C$ from Eq.~(\ref{ec}).
The ratio, $\zeta$, of the isomer and isotope shifts 
\begin{equation}
\zeta=\frac{\langle r^2_{229m} \rangle -\langle r^2_{229}\rangle}{\langle r^2_{232} \rangle -\langle r^2_{229} \rangle}=0.035(4)
\end{equation}
was measured in~\cite{Thielking18} for two transitions in Th$^{2+}$. New isotope shift measurements and more advanced theoretical calculations~\cite{Safronova18} yielded an improved value $\langle r^2_{232} \rangle -\langle r^2_{229} \rangle =0.299(15)$\,fm$^2$, reducing the uncertainty in this value by a factor of 3, from 15\,\% to 5\,\%.   As a result, the uncertainty in the  $\Delta \langle r^2_{229} \rangle = \langle r^2_{229m} \rangle -\langle r^2_{229} \rangle =0.0105(13)$\,fm$^2$ was reduced from 17\% to  12\,\%~\cite{Safronova18}. 
We note that this improvement in the  accuracy is important for the determination of the $\alpha$-sensitivity $K$ due to a strong cancellation of the $\langle r^2 \rangle$ and $Q_0$ contribution, as discussed below.

Using the spectroscopic quadrupole moment for the ground state $\Qlab=3.11(2)$\,eb~\cite{2021Th3} and the ratio of the intrinsic quadrupole moments $Q_0^\mathrm{m}/Q_0 = 1.01791(2)$, we find 
$Q_0^\mathrm{m}=9.85(6) \:$fm$^2$ and $\Qlab^\mathrm{m}=1.77(1)$\,eb
\footnote{Note that we follow $Q_0$ designations of~\cite{Faddeev20}, which differs from~\cite{Thielking18}, where the value of with $Q_0$ was defined without the factor $q_e Z$. Using designations and units of~\cite{Thielking18}, we get $Q_0^\mathrm{m}=8.86 (6)\: e$b.}.

The measured value of $\Delta{Q_0}/Q_0=0.01791(2)$, where $\Delta{Q_0}=Q_0^\mathrm{m}-Q_0$, differs by a factor of 2.4 from the prediction of $\Delta{Q_0}/{Q_0}=0.0075(20)$ obtained assuming a constant nuclear volume for the ground and the isomeric state~\cite{Faddeev20}. Using this value, the authors arrived at a negative $K=-8200(2500)$~\cite{Faddeev20}.

We computed the respective volumes using
\begin{equation}
\label{eq:21}
V=\frac{4\pi}{3}R_0^3=\frac{4\pi}{3}a^2c
\end{equation}
and find a small but non-zero -0.055\,\% difference, with the isomer volume being smaller. This change in the volume is well within the 0.8\,\% experimental uncertainty. However, it is interesting to note that volume changes due to the change in the rms radius and $Q_0$ contribute with opposite signs, +0.05\,\% and -0.105\,\%, respectively, leading to some cancellation of the overall effect. 
We tested the dependence of $\Delta{Q_0}/{Q_0}$ on the volume change and find that even a small -0.055\,\% change leads to a difference by over a factor of two in the resulting $\Delta{Q_0}/{Q_0}$, making the
constant volume approximation unreliable for extracting this quantity from the difference in the $\langle r^2 \rangle$. 

Substituting experimental values into the expression for the Coulomb energy difference given by Eq.~(\ref{ec}) gives
\begin{eqnarray}
\Delta E_C&=&-0.154(19) \,\textrm{MeV} +0.203(4) \,\textrm{MeV} \\&=&
0.049(19) \,\textrm{MeV},
\end{eqnarray}
demonstrating strong cancellation between $\langle r^2 \rangle$ and $Q_0$ terms, making the $\Delta E_C$ and $K$ values extremely sensitive to $\Delta{Q_0}/Q_0$.
The enhancement factor is \begin{equation}
K=-18400(2300) + 24300(400) =5900(2300),
\end{equation}
where the uncertainty is overwhelmingly due to the uncertainty in $\Delta \langle r^2 \rangle$.
An improved measurement of $\zeta$ is needed for further improvement in the accuracy of $K$.

Such a high value of $K$ drastically increases the sensitivity of a nuclear clock to dark matter and other related new physics searches in comparison with current optical atomic clocks. The highest sensitivity in atomic clocks is $K=-6$ (for Yb E3 transitions~\cite{Yb}). Using a nuclear clock thus gives three orders of magnitude improvement for the same clock accuracy. This improvement translates into being able to probe dark matter with three orders of magnitude smaller couplings.

\textit{Effect of the octupole deformation.}--
We point out that the discussion above does not include uncertainties connected with the present theoretical model, but only uncertainties of the experimentally determined quantities. In particular, a contribution of the currently unknown nuclear octupole deformation to the Coulomb energy $E_C$ is not accounted for. This effect was evaluated in~\cite{Faddeev20}, where the axially symmetric surface of the nucleus $r(\theta)$ was expanded as 
\begin{equation}
r(\theta)=R_s\left[ 1+\sum_{n=1}^{N} \left(\beta_n {Y_n^0}(\theta) \right)  \right], \label{r}
\end{equation}
where $\beta_n$ are  deformation parameters, $Y_n^0(\theta)$ are spherical harmonics, and  $R_s$ is defined by normalizing the volume to that of the spherical nucleus with equivalent sharp
spherical radius $R_0$.
The normalization condition can be expressed as $\int \rho_q(\bm{r}) d^3\bm{r} =1$, where in the present approximation of constant charge density, $\rho_q=1/V$ is the charge density divided by the total charge $q_eZ$ and $V=(4\pi/3)R^3_0$:
\begin{equation}
\frac{2 \pi}{3}\int_0^\pi r^3(\theta)\: \textrm{sin}(\theta)\: d\theta  =V.
\end{equation}

For a pear-shaped, axially symmetric nucleus with quadrupole and octupole deformation,  $N = 3$.
The coefficient $\beta_1$ is determined from the condition that the center of mass of the
shape is at the origin of the coordinate system, $\int_V \bm{r}  d^3\bm{r} =0$~\cite{review} .
For a pure quadrupole deformation (i.e. prolate spheroid), $N=2$ and $\beta_1=0$. The nuclear properties are related to the $\beta$ coefficients via 
\begin{eqnarray}
\langle r^2 \rangle &=& \int r^2 \rho_q(\bm{r}) d^3\bm{r}\\
Q_0 &=& 2\int r^2 P_2({\cos} \:\theta) \rho_q(\bm{r}) d^3\bm{r},\\
Q_{30} &=& 2\int r^3 P_3({\cos} \:\theta) \rho_q(\bm{r}) d^3\bm{r},\label{Q3}
\end{eqnarray}
where $P_2$ and $P_3$ are the Legendre polynomials,  and $Q_{30}$ is the intrinsic charge octupole moment~\cite{review}. 

The expression for the Coulomb energy~\cite{nucl}, Eq.~(6.47) 
 \begin{equation}
E_C=\frac{3q_e^2 Z^2}{5R_0}\left(1 - \frac{1}{4\pi} \beta_2^2 - \frac{5}{14\pi} \beta_3^2 \right), \label{qq}
\end{equation}
where $\beta_2$ and $\beta_3$ are the quadrupole and octupole deformations, respectively, and where we omit $O(\beta_n^3)$ terms.

The change in the Coulomb energy was described in~\cite{Faddeev20} using the expression 
\begin{eqnarray}
\Delta E_C &=&\frac{\partial E_C}{\partial \beta_2^2 } \Delta \beta_2^2 + \frac{\partial E_C}{\partial  \beta_3^2} \Delta \beta_3^2\\
&=&\frac{3q_e^2 Z^2}{5R_0}\left(- \frac{1}{4\pi}\Delta \beta_2^2 - \frac{5}{14\pi} \Delta\beta_3^2 \right). \label{dec}
\end{eqnarray}

However, this expression utilizes the constant volume approximation, while $R_0$ depends on $\beta_2$ and $\beta_3$
 via the volume normalization above. 
As in the ellipsoid model above, this leads to a wrong result for $\Delta E_C$, when using the experimental values for $\langle r^2 \rangle$ and $Q_0$.

 To evaluate the effect of a possible octupole deformation described by the change in $\beta_3$ upon nuclear excitation, we calculate
$\Delta E_C$ using experimentally available numbers. 
We use the definitions for $\langle r^2 \rangle$ and $Q_0$ and the normalization condition above, the experimental values for $\langle r^2 \rangle$ and $Q_0$,  to calculate $R_s$, $\beta_2$, and $R_0$ for both ground and isomeric state. 

First, we take $\beta_3=0$ for both the ground and isomeric state. The resulting values of $\beta_2=0.220$ and $\beta_2^{\rm{m}}=0.223$ 
are consistent with $^{228}$Th value from ~\cite{Th228}. 
We compute $\Delta E_C= 0.052$~MeV giving $K=6300(2300)$ using Eq.~(\ref{qq}), in agreement with the result obtained for the ellipsoid model (we note that Eq.~(\ref{qq}) is approximate). Then, we repeat the same computation but use $\beta_3=\beta_3^{\rm{m}}=0.115$ from~\cite{Palffy21}, which yields essentially the same result for $K$, as expected.
However, when we vary $\beta_3$ differentially between the ground and excited state, while keeping all other parameters constant, we find that 1\,\%, 3\,\%, and 5\,\% differential change in $\beta_3$ lead to $\Delta K = 2850$, $\Delta K = 8600$, and $\Delta K = 14500$, respectively. A larger  $\beta_3$ of the isomer yields a positive change in $K$. 

For reference, the measured change in $Q_0$, and correspondingly $\beta_2$, is 1.8\,\%. This computation shows that even 1\,\% variation of the octupole deformation changes $K$ by more than $1 \sigma$. Therefore, it is important to estimate the sign and constrain the magnitude of the octupole deformation change between the ground and isomeric state. 


We can estimate the approximate size of the intrinsic charge octupole moment in $^{229}$Th ground state and isomer. Using the $\beta_2$ value obtained above and $\beta_3$ within the range used in~\cite{Palffy21}, 0.11 to 0.145, we get 
$Q_{30}=35-44$~fm$^3$ [Eq.~(\ref{Q3})]. This is consistent with estimates~\cite{Th228} for the octupole moment of $^{228}$Th~\footnote{Note that~\cite{Th228} gives the values of $Zq_eQ$.}. E3 matrix elements of nuclear Coulomb excitation can be measured to determine the octupole moment, as has been demonstrated~\cite{gaffney2013studies}. We also note that even higher moments might contribute to the nuclear Coulomb energies. Octupole deformation in $^{229}$Th allows for the search of permanent electric-dipole moments~\cite{Th228,EDM}.




\textit{Conclusion.}-- 
The precise measurement of the change in nuclear quadrupole moment $\Delta Q_0/Q_0$~\cite{Zhang24} between the $^{229}$Th ground and isomeric state together with previous measurements of the mean-square charge radius $\langle r^2 \rangle$ allow us to extract the change in Coulomb contribution $\Delta E_C$ to the nuclear energies. We quantify the sensitivity of the nuclear transition to changes of the fine-structure constant $\alpha$, with the factor $K=5900(2300)$. Future theoretical and experimental work will focus on determining the octupole contribution to the nuclear energies, as we have shown that small octupole deformations strongly affect the alpha sensitivity. In addition, an improved measurement of the ratio $\zeta$ of the isomer and isotope shifts is required. The quantitative deviation from the constant volume approximation illustrates the power of the emerging precision laser spectroscopy on $^{229}$Th for testing fundamental models in nuclear physics.

\textit{Acknowledgements.}-- 
We thank Victor Flambaum, David DeMille, Julian Berengut, Adriana Palffy, Nikolay Minkov, Ronald Fernando Garcia Ruiz, Martin Pimon, Andreas Grüneis, Logan Hillbery, and Dahyeon Lee for stimulating discussions. 

This work has been funded by the European Research Council (ERC) under the European Union’s Horizon 2020 research and innovation programme (Grant Agreement No. 856415) and the Austrian Science Fund (FWF) [Grant DOI: 10.55776/F1004, 10.55776/J4834, 10.55776/ PIN9526523]. The project 23FUN03 HIOC [Grant DOI: 10.13039/100019599] has received funding from the European Partnership on Metrology, co-financed from the European Union’s Horizon Europe Research and Innovation Program and by the Participating States. K.B. acknowledges support from the Schweizerischer Nationalfonds (SNF), fund 514788 “Wavefunction engineering for controlled nuclear decays".

We also acknowledge funding support from the Army Research Office (W911NF2010182); Air Force Office of Scientific Research (FA9550-19-1-0148); National Science Foundation QLCI OMA-2016244; National Science Foundation PHY-2317149; and National Institute of Standards and Technology.

\bibliography{MS}
\bibliographystyle{ieeetr}

\end{document}



\title{Alpha-sensitivity of the Th-229 nuclear clock transition: Supplemental Materials}

\author{Kjeld Beeks$^{1,2}$}
\author{Georgy A. Kazakov$^{1}$}
\author{Fabian Schaden$^{1}$}
\author{Ira Morawetz$^{1}$}
\author{Luca Toscani De Col$^{1}$}
\author{Thomas Riebner$^{1}$}
\author{Michael Bartokos$^{1}$}
\author{Tomas Sikorsky$^{1}$}
\author{Thorsten Schumm$^{1}$}
\email{email: thorsten.schumm@tuwien.ac.at}
\affiliation{$^1$Vienna Center for Quantum Science and Technology, Atominstitut, TU Wien, 1020 Vienna, Austria}
\affiliation{$^2$Laboratory for Ultrafast Microscopy and Electron Scattering (LUMES), Institute of Physics, École Polytechnique Fédérale de Lausanne (EPFL), Lausanne CH-1015, Switzerland}

\author{Chuankun Zhang, Tian Ooi, Jacob S. Higgins, Jack F. Doyle, Jun Ye}
\affiliation{JILA, NIST and University of Colorado, Department of Physics, University of Colorado, Boulder, CO 80309}

\author{Marianna S. Safronova}
\affiliation{Department of Physics and Astronomy, University of Delaware, Newark, Delaware 19716, USA}
\date{\today}
\maketitle
\textit{Quadrupole structure.}--
In the nuclear laser spectroscopy performed in~\cite{Zhang24}, a sample containing $^{229}$Th nuclei doped in a CaF$_2$ matrix was studied~\cite{beeks2022nuclear,beeks2023growth,beeks2024optical}.
The interaction of the nuclear quadrupole moment $Q$ with the crystal electric field gradient (EFG) $V_{ij}$ leads to the emergence of a nuclear quadrupole splitting of the isomer transition~\cite{Kazakov12}. This structure allows us to determine the ratio between the quadrupole moments of the ${\rm ^{229}Th}$ nucleus in the ground and the isomeric state, as well as the local EFG. Here we describe this routine in more detail.

Assuming that the nuclear deformation is axially symmetric, and the nuclear spin has a well-defined direction with respect to the symmetry axis~\cite{moment}, the {\em spectroscopic} nuclear quadrupole moment $\Qlab$ is related to the {\em intrinsic} nuclear quadrupole moment $Q_0$ via the relation
\begin{equation}
Q_0=\frac{\Qlab}{q_e Z}\frac{(2I+3)(I+1)}{3k^2-I(I+1)},
\label{eq:QviaQ0}
\end{equation}
where $Z=90$ is the charge number of the thorium nucleus, $q_e$ is elementary charge,  $I$ is the nuclear spin ($I=5/2$ for ground, and $I=3/2$ for isomeric state), and $k$ is the projection of the nuclear spin on the deformation axis. Here we follow $\Qlab$ designation of \cite{Thielking18, moment} for spectroscopic, and $Q_0$ designation from \cite{Faddeev20} for intrinsic quadrupole moments, this is why an extra factor of $q_e Z$ appears in expression of $\Qlab$ via $Q_0$. $\Qlab$ is expressed in units of eb, and $Q_0$ in units of $\rm fm^2$ throughout this paper. The nuclear ground and isomeric states are the band heads of their respective rotational bands and therefore $k=I$~\cite{neugart2006nuclear}.

The EFG at the position of the nucleus can be expressed via the local electrostatic potential $V$ of the crystal lattice environment as
\begin{equation}
V_{ij}=\frac{\partial^2V}{\partial x_i \partial x_j}.
\label{eq:2}
\end{equation}
Let us consider the EFG in the ``main axes'', such that $V_{ij}$ is diagonal, and $|V_{zz}|\geq|V_{xx}|\geq|V_{yy}|$. Then we introduce the asymmetry parameter 
\begin{equation}
\eta = \frac{V_{xx}-V_{yy}}{V_{zz}},
\label{eq:4}
\end{equation}
and note that $0\leq \eta \leq 1$.

We can neglect the contribution of the local electron density at the position of the nucleus into the EFG. Such a contribution is responsible for the isomer shift, which is not considered here. Then, the EFG becomes a traceless tensor, and the Hamiltonian describing the quadrupole interaction can be written as \cite{Greenwood71}
\begin{multline}
{\HH_{Q}=} { \frac{V_{zz}}{4} 
\left(\frac{\Qlab\PP_g}{I(2I-1)}+\frac{\Qlab^m\PP_i}{I'(2I'-1)} 
\right) \times} \\
{\left\{ 3\I_z^2-\I^2+\eta(\I_x^2-\I_y^2) \right\}
}
\label{eq:5}
\end{multline}
where $\PP_{g,i}$ are projection operators to all the sublevels of the ground and the isomeric states, $\Qlab$ and $\Qlab^m$ are the spectroscopic quadrupole momenta of these states, respectively.

Energy levels and eigenstates $\ket{g}$, $\ket{i}$ can be obtained from diagonalization of the Hamiltonian (\ref{eq:5}). Then the eigenstates $\ket{g}$, $\ket{i}$ can be written as
\begin{align}
\ket{i}&=\sum_{m'}\alpha_{im'}\ket{I'm'}, ~~~~~~
\ket{g}&=\sum_{m}\beta_{gm}\ket{Im},
\label{eq:7}
\end{align}
where $\ket{Im}$ ($\ket{I'm'}$) are sublevels of the ground (isomeric) nuclear state with projection $m$ ($m'$) of nuclear moment $I$($I'$) on the main axis $z$ of the EFG and $\alpha_{im'}$ and $\beta\beta_{gm}$ the coefficients of expansion. 

Note that although for $\eta \neq 0$ the Hamiltonian (\ref{eq:5}) does not commute with the operator $\I_z$ and eigenstates of this Hamiltonian are not the states $\ket{I,m}$ with a defined projection $m$, one specific component $\ket{Im}$ or $\ket{I'm'}$ dominates in each of the eigenstates $\ket{g}$ or $\ket{i}$. In particular, $|\bra{g}\ket{Im}|>0.88$ and $|\bra{i}\ket{I'm'}|>0.96$ for such components up to $\eta=1$. Therefore, one may approximately characterise the states $\ket{g}$ ($\ket{i}$) by the values of $m$ ($m'$) corresponding to the dominating component.

\textit{Excitation probabilities.}--
The interaction Hamiltonian $\hat{U}$ of the nucleus with a laser field $\vec{B}=\vec{B}_0 \cos(\omega t)$ can be written (in a properly chosen rotating frame) as 
\begin{equation}
\hat{U}=\frac{1}{2}\sum_{i,g,q}\ket{i}B_0^q\bra{i}\hat{\mu}_{1,q}\ket{g}\bra{g} + h.c.,
\label{eq:8}
\end{equation}
where $B_0^q=|\vec{B}_0| \epsilon^q$ is the $q$th cyclic projection of the amplitude of the magnetic field of the laser, $\vec{\epsilon}$ is the polarization unit vector of this field, $\mu_{1,q}$ is the $q$th cyclic projection of magnetic moment operator $\hat{\mu}$. From the Wigner-Eckart theorem \cite{VMK} we can write
\begin{equation}
\begin{split}
\bra{i}\mu_{1,q}\ket{g} & 
=\sum_{m',m}\bra{I'm'}\mu_1^q\ket{Im} \alpha_{im'}^* \beta_{gm} \\ 
&=\frac{||\mu||}{\sqrt{2I'+1}}\sum_{m',m} C_{Im1q}^{I'm'} \alpha_{im'}^* \beta_{gm}, 
\end{split}
\label{eq:9}
\end{equation}
therefore, the transition matrix elements can be written as
\begin{equation}
\begin{split}
U_{ig} = \frac{||\mu|||\vec{B}_0|}{\sqrt{2I'+1}}\sum_{m',m,q} \epsilon_q C_{Im1q}^{I'm'} \alpha_{im'}^* \beta_{gm}. 
\end{split}
\label{eq:10}
\end{equation}
The excitation rate for any transition $\ket{g}\rightarrow \ket{i}$ is proportional to $|U_{ig}|^2$. To obtain relative intensities of the measured transition lines, we sum $|U_{ig}|^2$ over the degenerate states $\ket{g}$ and $\ket{i}$ with the same energies $E_1$ and $E_2$, and average over 3 possible orientations of the main axes of the EFG with respect to a polarization vector $\vec{\epsilon}$.
After some algebra, we find the intensity of the line with energy $E_2-E_1$ to be proportional to
\begin{equation}
\begin{split}
W(E_2-E_1) = \frac{1}{4}
\sum_{g,i}
\sum_{q=-1}^{1}\left| \sum_{m',m} C_{Im1q}^{I'm'} \alpha_{im'}^* \beta_{gm} \right|^2, 
\end{split}
\label{eq:12}
\end{equation}
where the first sum is taken over such states $i,g$, that their energies are $E(g)=E_1$ and $E(i)=E_2$.


\textit{Data Fitting.}--
In~\cite{Zhang24}, an absolute frequency measurements of five lines was performed, which were identified as transitions between certain quadrupole sublevels of the ground and the isomeric state of the $^{229}$Th nucleus, see Table~I of the main text. The experimental line intensities qualitatively agree with the theoretical expectations.

The high quality of the spectroscopy data allows to separate the two material parameters $V_{zz}$ and $\eta$, specific to the CaF$_2$ crystal environment (and common to ground and isomeric state), from the fundamental $^{229}$Th nuclear parameters $Q_0^m/Q_0$ and $\nu_{\rm Th}$. The transition frequencies are arithmetically consistent within few kHz for measurements performed over $>$10 days; we therefore assume fluctuations $V_{ij}$, i.e. due to changes in temperature or pressure, to be negligible.

We consider a 4-parametric model, where the 5 measured frequencies $\nu_i$ are fitted by 5 functions $f_j(\theta)$ depending on 4 fit parameters: $\theta_1=\nu_{\rm Th}$, $\theta_2=Q_0^m/Q_0$, $\theta_3=\Qlab V_{zz}$ and $\theta_4=\eta$. The fitted values $f_j(\theta)$ of the observed frequencies $\nu_i$ can be represented as
\begin{equation}
    f_{j}(\theta)=\theta_1+\frac{E_{2,j}-E_{1,j}}{2\pi \hbar},
    \label{eq:fit}
\end{equation}
where the energies $E_{2,j}$ and $E_{1,j}$ are obtained from diagonalization of the Hamiltonian~(\ref{eq:5}) taken with parameters $\theta_2,\theta_3$ and $\theta_4$, and corresponds to the ground ($E_{1,j}$) and the isomer ($E_{2,j}$) state with asymptotic magnetic quantum numbers $m_g$ and $m_e$ given in $j$th row of the Table~I of the main text. Then we introduce the normalized total variance $q$ characterising the goodness of the fit
\begin{equation}
    q(\theta)=\sum_{j=1}^5 \frac{(f_{j}(\theta)-\nu_j)^2}{\sigma^2_j},
    \label{eq:fitQual}
\end{equation}
where $\sigma_j$ is the experimental uncertainty of the frequency $\nu_j$ ($\sigma_1=\sigma_2=\sigma_3=\sigma_4=4~{\rm kHz}$, 
$\sigma_5=20~{\rm kHz}$). We find that the minimal value $q^L=0.4386$ is attained at 
$\nu_{\rm Th}=\theta_1^L=2020\,407\,384\, 335.1795~{\rm kHz}$, 
$Q_0^m/Q_0=\theta_2^L=1.0179120716$, $\eta=\theta_3^L=0.591636416$ and
$\Qlab V_{zz}=\theta_{4}^L=339.2634098~{\rm eb\, V/\r{A}^2}$; these values are forming the {\em least square estimator} $\theta^L$ of parameters $\theta$. If our 4-parametric model is correct (i.e., all 5 measured lines are elements of a quadrupole structure resulting from a single type of charge compensation configuration), $q^L$ should be a $\chi^2_{1}$ random value. Because the probability to find a $\chi^2_1$- random value below $q^L$ is about 0.492,  we can accept the 4-parametric model (\ref{eq:fit}) as an adequate description of the experimental data.

Now we can determine the covariance matrix ${\rm Var}[\theta^L]$ of the least square estimator $\theta^L$. According to \cite{BatesWatts}, this matrix can be calculated as
\begin{equation}
    {\rm cov}(\theta^L_r,\theta^L_s)=[(\X^T\X)^{-1}]_{rs}, \quad {\rm where}
    \quad 
    \X_{jr}=\frac{1}{\sigma_{j}}\frac{\partial f_j}{\partial \theta_r}.
    \label{eq:CovMatr}
\end{equation}
Square roots of diagonal elements of this covariance matrix are deviations $\varsigma_k$ of parameters $\theta_k$, particularly, we find $\varsigma_1=2.1$\,kHz, $\varsigma_2=1.7\times 10^{-5}$, $\varsigma_3=4.8\times 10^{-5}$ and $\varsigma_{4}=6.9\times 10^{-3}$~{eb\,V/\r{A}\textsuperscript{2}. Therefore, estimations of the fit parameters reads as: $\nu_{\rm Th}=2020\,407\,384\, 335(2)~{\rm kHz}$, 
$Q_0^m/Q_0=1.01791(2)$, $\eta=0.59163(5)$ and
$\Qlab V_{zz}=339.263(7)$~eb\,V/\r{A}\textsuperscript{2}, where all values are rounded to single-digit uncertainties.

We also present the correlation matrix
\begin{equation}
 \frac{{\rm cov}(\theta^L_r,\theta^L_s)}{\varsigma_r \varsigma_s}=
    \left(
    \begin{array}{cccc}
    1     & 0.018 & 0.30  & -0.17 \\
    0.018 &   1   & 0.025 & -0.25 \\
    0.30  & 0.025 &  1    & -0.34 \\
    -0.17 & -0.25 & -0.34 &  1
    \end{array}
    \right)_{rs}.
\label{eq:CorrMatrValues}
\end{equation}


\textit{Confidence regions.}--
Due to the pivotal relevance of the quadrupole moment ratio for estimating the change in Coulomb energy $\Delta E_C$, we explicitly constructed the confidence region for the two nuclear parameters $Q_0^m/Q_0$ and $\nu_{\rm Th}$. To do it, we performed a comparison of the minimal normalized variance $q^L$ for the {\em long model}~(\ref{eq:fit}), containing $l=4$ fit parameters $\theta_j$, with a minimal normalized variances $q^S(\theta_1,\theta_2)$ for a series of {\em short models}, where parameters $\theta_1=\nu_{\rm Th}$ and $\theta_{2}=Q_0^m/Q_0$ are fixed while the other $s=2$ parameters are fitted. Then, if the observed frequencies $\nu_j$ are normally distributed near their expectation values and the short model is correct, (i.e., if the parameters $(\theta_1,\theta_2)$ are true), then the difference $q^S-q^L$ is a $\chi^2_{l-s}=\chi^2_2$-random value (with 2 degrees of freedom). Therefore, we can characterize a pair $(\theta_1,\theta_2)$ by the confidence level $c$ defined as the probability that the $\chi^2_{l-s}$-random value is larger than $q^S-q^L$:
\begin{equation}
c(\theta_1,\theta_2)=P[\chi^2_2>q^S(\theta_1,\theta_2)-q^L]. 
\label{eq:ConfLevel}
\end{equation}
\begin{figure}
    \centering
    \includegraphics[width=0.35\textwidth]{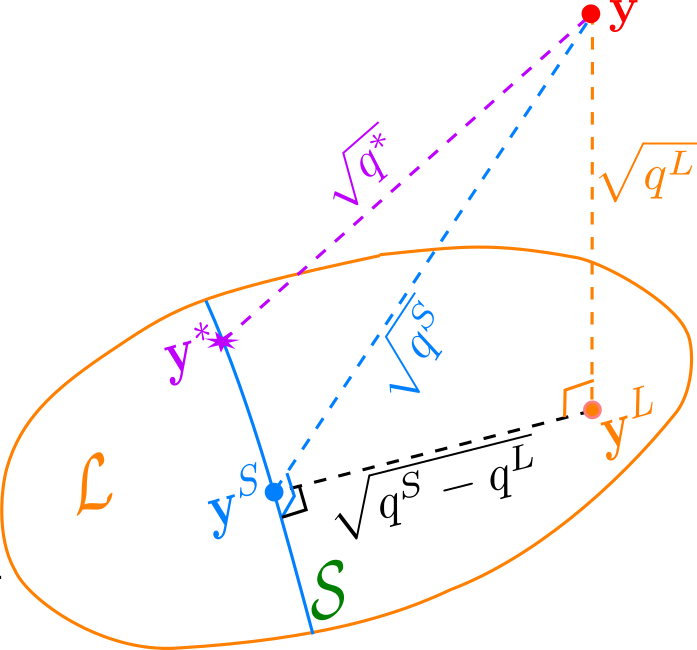}
    \caption{Sketch of the sample space. Here $\mathcal{L}$ and $\mathcal{S}$ are the long and the short hypothesis surfaces respectively. The point $\mathbf{y}^*=\mathbf{f}(\theta^*)$  corresponds to the true value of parameters. Points $\mathbf{y}^L=\mathbf{f}(\theta^L)$ and $\mathbf{y}^S=\mathbf{f}(\theta^S)$ corresponds to the best fits within the short and the long hypotheses respectively. Adopted from \cite{Kazakov14Heidelberg}.}
    \label{fig:SampleSpace}
\end{figure}

To illustrate the essence of these characteristics, let us represent our $N=5$ observations $y_j=\nu_j/\sigma_j$ as a point $\mathbf{y}$ in $N$-dimensional Euclidean {\em sample space}, see Figure~{\ref{fig:SampleSpace}}. In the {\em long model} it is supposed that the expectation values of observations $y_j$ are the known functions $f_j(\theta)/\sigma_j$ of the $l$-dimensional parameter $\theta=\{\theta_1,...,\theta_l\}$ ($l=4$ in our case). 
These functions form a $l$-dimensional hypersurface $\mathcal{L}$ in our sample space, and the distance between $\mathbf{y}$ and $\mathcal{L}$ is equal to $\sqrt{q^L}$. Similarly, the {\em short model} is a particular case of the long model, with additional restriction on parameters $\theta$ (in our case, a pair of these parameters is fixed), therefore, it has $s<l$ degrees of freedom ($s=2$ in our case), and the functions $f_j(\theta)/\sigma_j$ with restricted $\theta$ forms a $s$-dimensional hypersurface $\mathcal{S}$ lying within $\mathcal{L}$.

Let us suppose that both the long and the short models are true, i.e., that the point representing expectation values $\mathbf{y}^*=\mathbf{f}(\theta^*)$ lies in $\mathcal{S}$ (and, therefore, in $\mathcal{L}$). For normally distributed observations $y_j$ the distance $|\mathbf{y}-\mathbf{y}^*|$ is the square root of a $\chi^2_{N}$-random value $q^*$. Neglecting curvatures of the hypersurfaces $\mathcal{L}$ and $\mathcal{S}$ we see that the squares of distances $|\mathbf{y}-\mathbf{y}^L|^2$ and $|\mathbf{y}-\mathbf{y}^S|^2$ between the point $\mathbf{y}$ and hypersurfaces $\mathcal{L}$ and $\mathcal{S}$ are $\chi^2_{N-l}$ and $\chi^2_{N-s}$-random values respectively. In turn, $|\mathbf{y}^S-\mathbf{y}^L|^2$ is a $\chi^2_{l-s}$-random value. Supposing that the long model is true, and considering a series of short models with different values of fixed parameters, we can characterize the plausibility of every such a short model with the help of a {\em confidence level} $c$ defined as the probability that a $\chi^2_{l-s}$-random value is larger than $q^S-q^L$, according to~(\ref{eq:ConfLevel}).

The map of confidence levels for the pairs of ``nuclear'' parameters $(\nu_{\rm Th}, Q_{0}^m/Q_0)$ are shown in Figure~\ref{fig:ConfReg}~(a). In Figure~\ref{fig:ConfReg}~(b) we present a similar map for ``material'' parameters $(\Qlab\cdot V_{zz},\eta)$.
\begin{figure*}
    \centering
    \includegraphics[width=\textwidth]{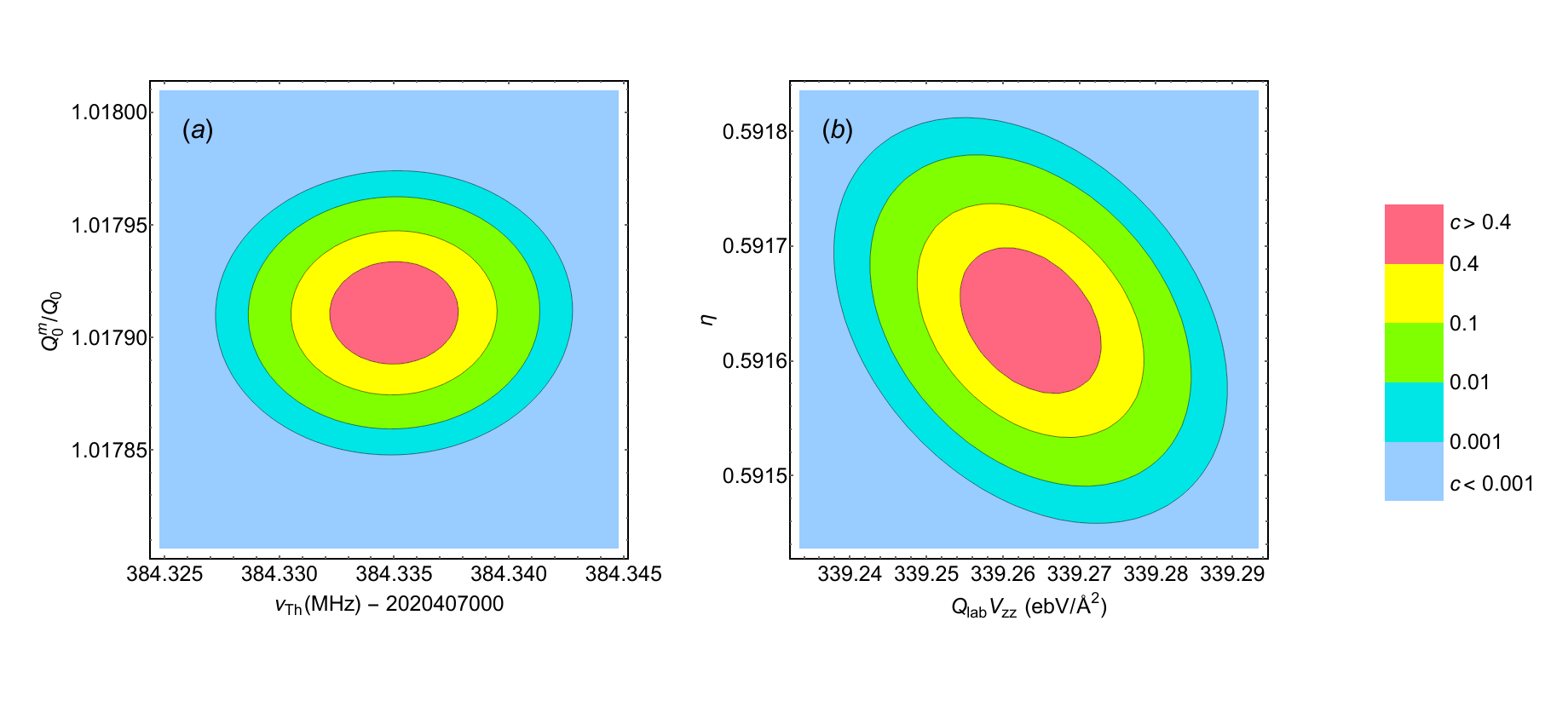}
    \caption{Confidence regions for ``nuclear'' (a) and ``material'' (b) parameters. 
    }
    \label{fig:ConfReg}
\end{figure*}

\bibliography{MS}
\bibliographystyle{ieeetr}